\documentclass[12pt]{article}
\usepackage{amssymb,epsf}
\def\lsim{\mathrel{\rlap {\raise.5ex\hbox{$ < $}}
{\lower.5ex\hbox{$\sim$}}}}
\def\gsim{\mathrel{\rlap {\raise.5ex\hbox{$ > $}}
{\lower.5ex\hbox{$\sim$}}}} 
\def\sqr#1#2{{\vcenter{\vbox{\hrule height.#2pt
        \hbox{\vrule width.#2pt height#1pt \kern#1pt
           \vrule width.#2pt}
        \hrule height.#2pt}}}}

\def\lsim{{\displaystyle
{{\raise-8pt\hbox{$ <$}}
\atop{\raise5pt\hbox{$\sim$}}}}}
\def\gsim{{\displaystyle
{{\raise-8pt\hbox{$ >$}}
\atop{\raise5pt\hbox{$\sim$}}}}}
\def\slsim{{\displaystyle
{{\raise-8pt\hbox{$\scriptstyle <$}}
\atop{\raise5pt\hbox{$\scriptstyle \sim$}}}}}
\def\sgsim{{\displaystyle
{{\raise-8pt\hbox{$\scriptstyle  >$}}
\atop{\raise5pt\hbox{$\scriptstyle \sim$}}}}}
\newskip\humongous \humongous=0pt plus 1000pt minus 1000pt

\newcommand{\sumpf}[0]{\sum_{(H^{\rm f},G^{\rm f})}\! \! \! \!
{\raise
4pt
\hbox{$'$}}\,}

\newcommand{\sump}[0]{\sum_{(H,G)}\! \! {\raise 4pt \hbox{$'$}}\,}

\def\bs{\begin{subequations}}
\def\es{\end{subequations}}
\catcode`\@=11
\newcount\hour
\newcount\minute
\newtoks\amorpm
\hour=\time\divide\hour by60
\minute=\time{\multiply\hour by60 \global\advance\minute by-\hour}
\edef\standardtime{{\ifnum\hour<12 \global\amorpm={am}%
        \else\global\amorpm={pm}\advance\hour by-12 \fi
        \ifnum\hour=0 \hour=12 \fi
        \number\hour:\ifnum\minute<10 0\fi\number\minute\the\amorpm}}
\edef\militarytime{\number\hour:\ifnum\minute<10 0\fi\number\minute}
\def\draftlabel#1{{\@bsphack\if@filesw {\let\thepage\relax
   \xdef\@gtempa{\write\@auxout{\string
      \newlabel{#1}{{\@currentlabel}{\thepage}}}}}\@gtempa
   \if@nobreak \ifvmode\nobreak\fi\fi\fi\@esphack}
        \gdef\@eqnlabel{#1}}
\def\@eqnlabel{}
\def\@vacuum{}
\def\draftmarginnote#1{\marginpar{\raggedright\scriptsize\tt#1}}
\def\draft{\oddsidemargin -.2truein
        \def\@oddfoot{\sl preliminary draft \hfil
        \rm\thepage\hfil\sl\today\quad\militarytime}
        \let\@evenfoot\@oddfoot \overfullrule 3pt
        \let\label=\draftlabel
        \let\marginnote=\draftmarginnote
   \def\@eqnnum{(\theequation)\rlap{\kern\marginparsep\tt\@eqnlabel}%
\global\let\@eqnlabel\@vacuum}  }
\def\subequations{\refstepcounter{equation}%
  \edef\@savedequation{\the\c@equation}%
  \@stequation=\expandafter{\theequation}
  \edef\@savedtheequation{\the\@stequation}
  \edef\oldtheequation{\theequation}%
  \setcounter{equation}{0}%
  \def\theequation{\oldtheequation\alph{equation}}}

\def\endsubequations{\setcounter{equation}{\@savedequation}%
  \@stequation=\expandafter{\@savedtheequation}%
  \edef\theequation{\the\@stequation}\global\@ignoretrue
  \vspace*{-12pt} \\}
\def\bs{\begin{subequations}}
\def\es{\end{subequations}}
\relax
\def\Im{\,{\rm Im}\, }

\def\thefootnote{\fnsymbol{footnote}}
\def\be{\begin{equation}}
\def\ee{\end{equation}}
\def\ba{\begin{eqnarray}}
\def\ea{\end{eqnarray}}

\def\ee{\end{equation}}
\def\bea{\begin{eqnarray}}
\def\eea{\end{eqnarray}}

\def\np#1#2#3{Nucl. Phys. {\bf{B#1}} (#2) #3}
\def\pl#1#2#3{Phys. Lett. {\bf{B#1}} (#2) #3}

\def\pr#1#2#3{Phys. Rev. {\bf{D#1}} (#2) #3}

\newcommand{\uarrw}[0]{\mathrel{
{\raise.5ex\vbox{\hrule width 1cm}\hskip-6pt\rightarrow}}}
\def\thebibliography#1{%
\vskip 0.5cm \centerline{\bf References}
\list{%
[\arabic{enumi}]}{\settowidth\labelwidth{[#1]}
\leftmargin\labelwidth
\advance\leftmargin\labelsep
\usecounter{enumi}}
\def\newblock{\hskip .11em plus .33em minus .07em}
\sloppy\clubpenalty4000\widowpenalty4000
\sfcode`\.=1000\relax}

\renewcommand{\theequation}{\arabic{section}.\arabic{equation}}

\renewcommand{\section}{\setcounter{equation}{0}\@startsection%
{section}{1}{0mm}{-\baselineskip}{0.5\baselineskip}%
{\normalfont\normalsize\bfseries}}

\renewcommand{\subsection}{\@startsection%
{subsection}{2}{0mm}{-\baselineskip}{0.5\baselineskip}%
{\normalfont\normalsize\slshape}}
\begin{document}
\renewcommand{\theequation}{\arabic{section}.\arabic{equation}}
\begin{titlepage}
\begin{flushright}
Bicocca-FT-00-08,\\
hep-th/0005198 
\end{flushright}
\begin{centering}
\vspace{1.0in}
\boldmath
{\bf \large Weak Scale in Heterotic String$^\dagger$}
\\
\unboldmath
\vspace{1.7 cm}
{\bf Andrea Gregori$^1$} \\
\medskip
\vspace{.4in}
{\it  Dipartimento di Fisica, Universit{\`a} di Milano--Bicocca},\\
{\it and}\\
{\it  INFN, Sezione di Milano, Italy}\\
\vspace{2.5cm}
{\bf Abstract}\\
\vspace{.1in}
We investigate the possibility of lowering the string scale
in four dimensional heterotic models possessing a non-perturbative extension 
of the gauge group. In particular, we consider a class of compactifications
in which the perturbative gauge sector is massive, and all the gauge bosons
are non-perturbative, with a coupling independent on the Planck and string 
scales.   
\end{centering}
\vspace{2cm}

\hrule width 6.7cm
$^\dagger$\  Research partially supported by the EEC under the contract\\
TMR-ERBFMRX-CT96-0045.\\
\\
$^1$e-mail: agregori@pcteor.mi.infn.it

\end{titlepage}
\newpage
\setcounter{footnote}{0}
\renewcommand{\thefootnote}{\arabic{footnote}}

\setcounter{section}{1}
\section*{\normalsize{\bf 1. Introduction}}

A problem of string theory is the apparent huge   
separation between the natural string mass scale and the 
scale of weak interactions. In the ${\cal N}_4=4$
heterotic string, at the tree level 
the string scale is related to the Planck scale by the value of the gauge
coupling, and a ``normal'' value of the latter implies that
its value is close to the Planck mass.
Unfortunately, this has the unappealing consequence that the string effects 
appear to be far from experimental detection.
Scenarios with a low compactification scale \cite{a}
or a low (or intermediate) string scale have therefore been considered
\cite{witten}--\cite{bo}. 
In the heterotic string, solutions to the problem of lowering the string scale
such as taking into account loop corrections or the effect
of the M-theory eleventh dimension \cite{witten} are characterized by 
the existence of bounds that prevent the string scale from being
very (arbitrarily) small.
In Ref. \cite{bo} it has been proposed that the problem 
of lowering the string scale, and having at the same time a 
value of the gauge coupling around $\sim 0,01$ or bigger, can be solved
by assuming that the gauge group corresponding to the Standard
Model belongs to a non-perturbative sector of the heterotic string.
The existence of an enhancement of the heterotic gauge group,
for a certain kind of compactifications, is a well known phenomenon:
it emerged in the context of string-string duality, when considering
the type I dual constructions \cite{sagn,gp}, in which this part of the gauge
group appears perturbatively.
On the heterotic side, the enhancement of the gauge group is explained
by the existence of instantons that shrink to zero size \cite{w}. 
The six dimensional coupling of this sector is therefore one,
and in four dimensions the coupling depends on the volume of the
further compactification, and not on the string and Planck scales.
Indeed, as we observed in Ref. \cite{gsmall},
in four dimensions the heterotic string possesses a further
non-perturbative extension of the gauge group, due to the
appearance, together with the small instantons of above, 
of gauge bosons deriving from six-dimensional massless tensors. 
The coupling of this sector does not depend
on the  volume of the compactification space, from six to four dimensions,
but rather on its shape. This gauge sector is also a candidate to accommodate 
the Standard Model group. 

The major problem related to this kind of  scenarios is however
the renormalization of the couplings, that in general leads to an
effective gauge coupling depending on all the original bare ones,
including that of the perturbative sector. This can eventually spoil
the independence of the gauge coupling on the string scale. 

In this note, we first discuss in some detail this problem.
We then consider phases of heterotic compactifications for which this does not
happen, being the gauge group  \emph{entirely} non-perturbative.

\noindent

\vskip 0.3cm
\setcounter{section}{2}
\setcounter{equation}{0}
\section*{\normalsize{\bf 2. Discussion}}

In four-dimensional compactifications of
the heterotic string, the tree level (perturbative) gauge coupling 
$\alpha_G$ is parametrized 
by the vacuum expectation value of the axion--dilaton field $S$:
\be
\alpha_G^{-1} \sim \Im S \, , ~~~~ 
\label{as}
\ee
where $S=a+{\rm i}{\rm e}^{- 2 \phi}$. 
In terms of string parameters, this reads:
\be
\alpha_G 
 \sim  { V_{(6)}  \over \lambda_H^{-2} l^6_H} \, ,
\ee
where $V_{(6)}$ is the volume of the six-dimensional compact space 
${\cal M}_6$, $l_H$ is the heterotic string scale
and $\lambda_H$ the heterotic coupling of ten dimensions.
Owing to the relation:
\be
G_N \equiv l_p^2 \sim { \lambda_H^2 l^8_H \over V_{(6)} } \, ,
\ee
there exist a tree level relation between the heterotic string scale,
gauge coupling, and the Planck mass:
\be
l_H \, = \, \langle \Im S  \rangle \, 
l_P \, = \, \alpha_{G_{(tree})}^{-1} \, l_P \, .
\label{lh}
\ee
According to this, the requirement of having a gauge coupling
of order $10^{-2} \div 10^{-1}$ seems to imply that the heterotic
string scale must be close to the Planck scale.
As discussed in Ref. \cite{bo}, the above statement is not anymore valid
if the ``Standard Model'' gauge group is provided by small
instantons, appearing in compactifications with
reduced supersymmetry, such as for ${\cal M}_6=T^2 \times K3$.
The coupling $\alpha_{G^{\prime}}$ of this sector 
depends on the volume of $T^2$:
\be
\alpha_{G^{\prime}}^{-1}  \sim  { V_{(2)} \over l^2_H} \label{50} \, ,
\ee 
($V_{(6)} \equiv V_{(2)} V_{K3}$), and is independent on the string
scale, still determined by (\ref{lh}).
As we discussed in Ref. \cite{gsmall}, T-duality of the heterotic string
implies that, together with the above, 
there is another non-perturbative gauge group,
with coupling given by the complex structure modulus of the torus:
\be
\alpha_{G^{\prime \prime}}^{-1}  \sim  { R_1 \over R_2 } \label{501} \, .
\ee 
This rather peculiar phenomenon tells us that, indeed,
in the heterotic string there is a huge freedom in the choice
of parameters, leading to a low string scale.
However, generically one loop corrections to the perturbative gauge coupling 
modify the expression  (\ref{as}):
\be
{1 \over \alpha_G} \approx \Im S + \beta \Delta (T) + 
\sum_i \beta_i  \Delta_i (U,Y) + f (T,U,Y) \, ,
\label{gst}
\ee
where $T$  and $U$ are the moduli related respectively
to the K\"{a}hler class and the complex structure of the two-torus
($\Im T \sim V_{(2)}= R_1 R_2$, $\Im U \sim R_1 \big/ R_2$), and
$\beta_i$, $\beta= \sum_i \beta_i$ are beta-function coefficients depending
on the specific gauge group factors, determined by the Wilson lines $Y$. 
Since the moduli $T$ and $U$ indeed parametrize the couplings of the
two non-perturbative extensions of the gauge group, we
interpret the dependence of $\alpha_G$ on these fields as due to the
running of states charged under all these sectors.
Based on this observation, we are led to conclude that
an analogous phenomenon should happen also in the non-perturbative sectors,
in which, generically, we expect that the gauge couplings
given in Eqs. (\ref{50}) and (\ref{501}) indeed acquire 
a dependence on the modulus $S$. Therefore,  
even in case the Standard Model gauge group had a non-perturbative origin,
there would be a relation between the gauge coupling and
string scale, leading, in the case of a non negligible 
dependence on the modulus $S$, to the loss of predictive power of the 
``tree level'' considerations.
The worse situation one can imagine is the one in which
perturbative and non-perturbative sectors behave in an analogous way,
with a simple exchange of the role of the fields $S$ and $T$ (or $U$).
In this case, since $\Delta(T) \sim \Im T$,
$\Delta(U,Y) \sim \Im U$ for large $\Im T$, $\Im U$, there would essentially
be no difference between perturbative and non-perturbative sector.
We notice that, among the very special situations in which
this mixing of couplings does not occur, there is the
case in which the gauge group is $U(16)$, namely the one for which
heterotic/type I duality has been tested in Ref.\cite{par}. 
The choice of such configurations seems however to put a too 
severe constraint on the allowed gauge groups.

It is however possible to escape this problem, by considering heterotic 
configurations in which the entire gauge group has a non-perturbative origin.
An indication that such configurations exist is given by
the expression of the correction to the coupling of the $R^2$ term
in an heterotic ${\cal N}_4=2$ models without perturbative 
gauge group \cite{sv,gkp2}, that read: 
\be
{1 \over g^2_{grav}} 
\sim  \Im S - \log \Im T | \vartheta_2 (T ) |^4 - 
\log \Im U | \vartheta_2 (U )|^4 \, ,
\ee
where for simplicity we don't specify normalization coefficients
and the term accounting for the infrared running 
(see Refs. \cite{gkp2}--\cite{gkp}). 
In the limit of large $\Im T$, $\Im U$, the second and third term
behave linearly in $\Im T$, $\Im U$, signaling the appearance of
new massless states in the corresponding non-perturbative sectors, 
of which these moduli parametrize the couplings. 
Indeed, there exists a type IIA/B self-mirror orbifold that
could describe the dual of this phase: it was constructed
in Ref. \cite{gkr}, 
as a semi-freely acting $Z_2 \times Z_2$ orbifold \cite{gm},
corresponding to a singular limit in the moduli space of the
compactification on the so-called Del Pezzo surface \cite{c}.
This CY$^{19,19}$ manifold is a double fibration over ${\bf P^1}$.
At the $Z_2$ orbifold point, this model
has two twisted sectors, each one providing eight vector multiplets and 
eight hyper multiplets.
According to Ref.~\cite{gkr}, the corrections to the $R^2$ term read:
\be
{1 \over g^2_{grav}} \sim - \log \Im T^1 | \vartheta_4 ( T^1)|^4 -
\log \Im T^2 | \eta (T^2)|^4 - \log \Im T^3 | \eta (T^3)|^4 \, ,  
\label{1616}
\ee    
where $T^1$, $T^2$, $T^3$ are the moduli associated to the
K\"{a}hler classes of the three tori into which the compact space is divided.
It is natural to identify the moduli $T^1$, $T^2$, $T^3$ respectively
with the heterotic fields $S$, $T$ and $U$ in the phase
in which non-perturbative massless sectors appear.
The theta function in the first term indicates in fact that
the corresponding sector, the perturbative sector of the heterotic string,
is massive, with a mass scaling roughly as $\sim \Im S$.
For large $\Im S$, the contribution of the first term 
diverges logarithmically, indicating, as usual \cite{solving}, 
the ``disappearance''
of the corresponding sector, or, in other terms, the fact that its
states are infinitely massive. We expect that this behavior
reflects in the gauge couplings of the non-perturbative sectors.
More precisely, since
in this ${\cal N}=2$ model the gauge group is realized at the level 2,
with an equal number of vector and hyper multiplets, and
the type II compactification manifold is self-mirror,
we don't expect corrections to the moduli spaces associated
to these states. There should be no corrections to the
gauge couplings either, that should be given by their ``bare'' value, 
as a function of the only field $T$ or $U$ respectively. However, even
in more realistic situations, such as those in which supersymmetry
is further broken to ${\cal N}=1$ or ${\cal N}=0$, 
there should not be strong corrections to the gauge couplings
depending on the field $S$, because the states of this sector are
infinitely massive. In this class of theories, therefore,
we expect the strength of the gauge coupling to be 
independent on the string scale, that can be arbitrarily low. 
 
In the opposite limit $\Im S \to 0$, namely in the S-dual situation, 
a better description is given in terms of $\tilde{S} \equiv -1 \big/ S$.  
The first term in Eq. (\ref{1616}) changes according to
$\vartheta_4(S) \to \vartheta_2(\tilde{S})$, and in the large
$\Im \tilde{S}$ limit the first term diverges linearly in $\Im \tilde{S}$,
indicating that the states of the corresponding sector, the
perturbative sector of the S-dual heterotic theory, are 
close to become massless ($m \sim 1 \big/ \tilde{S}$).
As discussed in Ref. \cite{gmonobis}, this model is in fact
probably connected to the ordinary ${\cal N}=2$ heterotic orbifold
with a rank 16 perturbative gauge group, and two 
equivalent non-perturbative sectors (cfr. Ref. \cite{gsmall}). This model is
dual to a $Z_2 \times Z_2$ non-freely acting type II orbifold.
The connection, at the limit $\Im \tilde{S} \to \infty$,
should however involve some kind of phase transition, in which
not only states of the ``$S$'' sector become massless, but also
in the two non-perturbative sectors new gauge bosons appear,
extending the rank of the gauge group from 8 to 16 in each of these  
sectors.

\noindent

\vskip 0.3cm
\setcounter{section}{3}
\setcounter{equation}{0}
\section*{\normalsize{\bf 3. Conclusions}}

In this paper we considered the problem of lowering the string scale 
in the heterotic string, without lowering the
gauge coupling, that should remain of the order $\sim 0,01$.
We found that indeed this is possible in a wide class of configurations.
This is made possible by  the fact that heterotic compactifications with
reduced supersymmetry possess several non-perturbative extensions
of the gauge group, for which the gauge coupling escapes the usual
tree level relation to the Planck and string scales.
An interesting class of compactifications is the one in which
the entire gauge group is non-perturbative.
Indeed, using type II/heterotic duality, it is possible to see
that, in the heterotic moduli space, perturbative and 
non-perturbative sectors are essentially equivalent \cite{gsmall}.
There is therefore no reason for preferring one sector among the others
to be the one giving origin to the Standard Model gauge group.
The latter could well lie on a non-perturbative sector
of the heterotic string, as proposed in Ref. \cite{bo}.

Our analysis was limited to ${\cal N}_4=2$ compactifications.
When supersymmetry is further broken to ${\cal N}_4=1$, we expect on general
grounds new non-perturbative sectors to appear, parallel
to the appearance of new moduli entering in the expressions of
string threshold corrections,
that we can associate to the couplings of these additional sectors.
This makes the scenario even reacher.  
On the other hand, non-perturbative phenomena have been shown
to play a crucial role in ${\cal N}_4=1$ heterotic compactifications,
in which most probably they are responsible for a further,
complete breaking of supersymmetry \cite{mayr}.

\vskip 1.cm
\centerline{\bf Acknowledgments}
\noindent
I thank I. Antoniadis, C. Bachas, K. Benakli, L. Girardello,
R. Minasian and B. Pioline
for valuable discussions, and the \'Ecole Normale of Paris for hospitality.


\end{document}